\documentclass[a4paper,12pt]{article}
\usepackage[body={16cm,24cm}]{geometry}
\usepackage{changepage}		% provides adjustwidth command used in definition 
\usepackage{eucal,latexsym,graphicx,amsfonts,amssymb,amsmath,mathrsfs}

\usepackage[labelfont=bf]{caption}
\usepackage[flushleft]{threeparttable}  % aligns table caption left
\usepackage{booktabs}   % needed for \toprule in tables
\usepackage[capposition=top]{floatrow}
\usepackage{bm}

\usepackage[justification=centering]{caption}
\usepackage{color}

\usepackage{harvard}
\usepackage{graphicx}

\PassOptionsToPackage{normalem}{ulem}
\usepackage{ulem}
\usepackage{array}
\usepackage{rotating}
\usepackage{units}
\usepackage{multirow}
\usepackage{lscape}
\usepackage{color,soul} %for highlighting
\usepackage{mathtools}
\usepackage{enumitem}  

\usepackage{amsthm}
\usepackage{amsmath}
\usepackage{amssymb}

  \theoremstyle{plain}
  
  \theoremstyle{plain}

  \makeatother

  \providecommand{\assumptionname}{Assumption}
  \providecommand{\propositionname}{Proposition}
\citationstyle{dcu}
\citationmode{default}

\makeatletter
\let\NAT@parse\undefined
\makeatother

\usepackage[colorlinks=true,linkcolor=blue,filecolor=magenta,urlcolor=cyan,hypertexnames=true]{hyperref}

\begin{document}
\begin{center}
\begin{Large}
\bf
Cryptocurrencies, Mainstream Asset Classes and Risk Factors -- A Study of Connectedness \\[6ex]
 \end{Large}

{\large  George Milunovich}\\
Macquarie University,
Sydney, NSW, 2109, Australia\\ 
george.milunovich@mq.edu.au \\[4ex]

\today\\[2ex]
\bigskip

\end{center}
\begin{abstract}
\noindent{}We investigate connectedness within and across two major groups or assets: i) five popular cryptocurrencies, and ii) six major asset classes plus two commonly employed risk factors. Granger-causality tests uncover six direct channels of causality from the elements of the mainstream assets/risk factors group to digital assets. On the other hand there are two statistically significant causal links going in the other direction. In order to provide some perspective on the magnitude of the uncovered linkages we supplement the analysis by estimating networks from forecast error variance decompositions. The estimated connectedness \textit{within} the groups is relatively large, whereas the linkages \textit{across} the two groups are small in comparison. Namely, less than 2.2 percent of future uncertainty of any cryptocurrency is sourced from all non-crypto assets combined, while the joint contribution of all digital assets to non-crypto uncertainty does not exceed 1.5 percent.

\noindent{}
 
\bigskip \bigskip 

\noindent {\em Key Words:} Bitcoin, Cryptocurrencies, Connectedness, Major Asset Classes, Risk Factors, Network, Granger Causality, Forecast Error Variance Decomposition
\\
{\em JEL classification:} C3, C32, G1

\end{abstract}

\newpage

\section{Introduction} \label{section_introduction}
"Virtual currencies, perhaps most notably Bitcoin, have captured the imagination of some, struck fear among others, and confused the heck out of the rest of us." \\
\mbox{}\hfill  -- Thomas Carper, US-Senator 
\bigskip \\

Cryptocurrencies are digital assets intended to serve as alternative means of payment. They are created and managed via decentralized open source code, rather than authority such as a central bank. Bitcoin, one of the most popular cryptocurrencies, was introduced in a whitepaper written by \citeasnoun{nakamoto2008bitcoin}. The whitepaper states that Bitcoin is a "peer-to-peer version of electronic cash [which] would allow online payments to be sent directly from one party to another without going through a financial institution". While the true identity of Nakamoto remains unknown it appears that Bitcoin was, at least in part, inspired by the 2008 Financial Crisis. This can be inferred from the text embedded into the first block of Bitcoins which reads: "The Times 03/Jan/2009 Chancellor on brink of second bailout for banks". While conceived in 2008, the Bitcoin network was developed in 2009 and 2010 recorded several initial transactions. 

More recently the growth in the number of Bitcoin transactions and its market cap has been unprecedented. In response to the success of Bitcoin new cryptocurrencies started entering the market in the early 2010. Presently there are more than 130 cryptocurrencies that have market capitalization in excess of USD 100 million. Given that high growth rates have arguably been the result of speculation, the rise of cryptocurrencies has led to significant confusion amongst investors, policy makers, financial institutions and general public.

Empirically Bitcoin is found to exhibit high levels of risk, see e.g. \citeasnoun{katsiampa2017volatility}, \citeasnoun{blau2017price} and \citeasnoun{pieters2017financial}, to be susceptible to the formation of speculative bubbles \cite{cheah2015speculative}, and to be prone to sudden crashes \cite{fry2016negative}. Interestingly, however, although the market has been in existence for less than a decade it has been found to be informationally efficient as reported in \citeasnoun{urquhart2016inefficiency}, \citeasnoun{nadarajah2017inefficiency} and \citeasnoun{bariviera2017some}. With reputable economists arguing both in favour of and against\footnote{For example, in a discussion of monetary policy IMF's He lists a number of advantages of crypto assets over fiat currencies \cite{he2018monetary}. On the other hand, an article titled "Stiglitz, Roubini and Rogoff lead joint attack on bitcoin" was recently published in financial press \cite{newlands2018}.} Bitcoin, as well as cryptocurrencies in general, it remains unclear how to value digital assets. 

In this paper we do not directly address the issue of valuing cryptocurrencies, but instead examine their time-series links with mainstream asset classes. Such measures of connectedness play a pivotal role in modern finance and are crucial for a number of reasons. As described in Section \ref{section_data} below, cryptocurrencies have displayed incredible price swings since their inception ten years ago. Therefore a pertinent question one may ask is whether digital asset prices can exert any influence on mainstream asset classes thereby destabilizing the entire financial system. This type of risk is often termed systemic risk in the literature. As noted in \citeasnoun{billio2012econometric} systemic risk involves the financial \textit{system}, which is a collection of interconnected organisations through which iliquidity, insolvency and losses can rapidly transfer during periods of unfavourable financial conditions. It is thus of interest to investigate if, and how, digital asset price movements impact the rest of the system. Second, from a diversification perspective, one may wonder how insulated the movements of crypto assets are from major market trends. If digital currencies are not impacted by the overall market, then one may argue in favour of diversifying into digital assets -- especially during periods of financial downturns such as the 2008 Financial Crisis. Third, since digital assets are expected to serve as a medium of exchange, it is of interest to find out how they interact with the US dollar which is widely regraded as a de facto world currency. 

Our analysis is concerned with the connectedness within and across two broad groups of financial variables: i) five digital assets, and ii) six mainstream asset classes plus two risk factors. The group of cryptocurrencies consists of Bitcoin, Etherium, Lightcoin, Monero and Ripple. These are some of the most liquid digital assets with significant market capitalizations. Mainstream assets are represented by stocks, US government bonds, US Dollar, oil, gold and a broad commodity index. We also include the TED spread\footnote{TED spread is a commonly used measure of the level of credit risk in the economy.} and a world geopolitical risk index proposed in \citeasnoun{caldara2018measuring} in our analysis. We use daily data spanning the August 2015 -- April 2018 time period.  A related study of \citeasnoun{corbet2018exploring} considers similar issues within a smaller system of assets and over an earlier time period.

Two econometric methods are employed to capture different aspects of connectedness. First, a purely predictive aspect of connectedness based on statistically significant lead-lag relationships is captured via a Granger-causality test, which are then mapped into a network as in \citeasnoun{billio2012econometric}. Second, relative magnitudes of connectedness are estimated by constructing a forecast error variance decomposition connectedness table proposed in \citeasnoun{diebold2014network}.

Our findings may be summarized as follows. Granger causality analysis uncovers connectedness both within and across the two groups of assets. We report six Granger causal relations from non-digital assets to cryptocurrencies, and two causal links going from digital to non-digital assets. While these uncovered relations are statistically significant, and therefore unlikely to be due to random chance, it is difficult to judge their magnitude on the basis of statistical tests alone. Thus we supplement our results with evidence from variance decomposition analysis. According to the measures of connectedness computed from variance decompositions within group links are relatively large, at times exceeding 15 percent. For instance, Bitcoin appears to have relatively high connectedness with three other digital assets. However, across group links appear to be rather negligible in magnitude. We find that combined contribution of all cryptocurrencies to the future uncertainty of non-digital variables does not exceed 1.5 percent, while less than 2.2 percent of the uncertainty of any cryptocurrency is accounted jointly by all non-digital assets. 

This paper proceeds as follows. In Section 2 we introduce the digital asset class and provide some basic descriptive statistics. Section 3 defines and discusses the two measures of connectedness which are employed to quantify the links between the variables of interest. Section 4 presents our main empirical findings, and Section 5 concludes.

\section{Data} \label{section_data}
We analyse daily data on five digital assets, six mainstream assets and two risk factors. Our cryptocurrencies include Bitcoin (BTC), Etherium (ETH), Lightcoin (LTC), Monero (XMR) and Ripple (XRP), which are some of the most actively traded digital assets. Amongst the cryptocurrencies, BTC is the best known asset. It represents the first successful implementation of the peer-to-peer currency design that has no central authority. While providing anonymity the BTC market design lacks any protection mechanisms from regulatory bodies, as discussed in \citeasnoun{vandezande2017virtual}. The remaining four cryptocurrencies considered here share similar characteristics, with the differences being in technical details regarding transaction processing algorithms, rewards for processing transactions (mining) and transaction fees. In addition, Monero can be singled out as the most anonymous cryptocurrency, with all its payments and account balances being entirely hidden. It employs an obfuscated public ledger, meaning that anybody can send transactions but no outside observer can tell the source, amount or destination of transacted funds. The combined market cap of the five digital assets at the time of writing is in excess of USD 195 billion\footnote{The market cap breakdown is roughly as follows: BTC -- \$129.8 bn, ETH -- \$42.4 bn, XPR -- \$16.9 bn, LTC -- \$4.5 bn and XMR -- \$1.9 bn. Source: \url{ https://coinmarketcap.com}.}.  

Our mainstream asset group comprises stocks -- S\&P 500 total return index (SP500), government debt as captured by the US benchmark 10 year government bond index (10YR), crude oil (WTI), gold (GOLD), commodities (SPGSCI) and the USD trade-weighted index. The two risk factors included in the analysis are the TED spread and a geopolitical risk index (GPR). The TED spread is the difference between the LIBOR rate and the short-term risk-free US Government debt. It is often used as a measure of credit risk. The GPR index is computed via automated text-search results of the electronic archives of eleven national and international newspapers. \citeasnoun{caldara2018measuring} calculate the index by counting the number of articles related to geopolitical tensions in each newspaper for each month (as a share of the total number of news articles). 

Our dataset consists of daily data collected over the 7 August 2015 -- 13 April 2018 time period. Digital asset price data is obtained from a cryptocurrency data hub CryptoCompare. The data for SP500, 10YR, WTI and SPGSCI data is from Datastream, while TED and USD are collected from the Federal Reserve Bank of St. Louis database FRED. Lastly, GPR index is obtained from \url{https://www2.bc.edu/matteo-iacoviello/gpr.htm}.

\begin{figure}[!h] 
\caption{Growth Rates in Asset Prices (computed as $\textrm{ln}[P_t/P_0]$)}\label{fig:time_series}
\includegraphics[height=4.5in,width=6in]{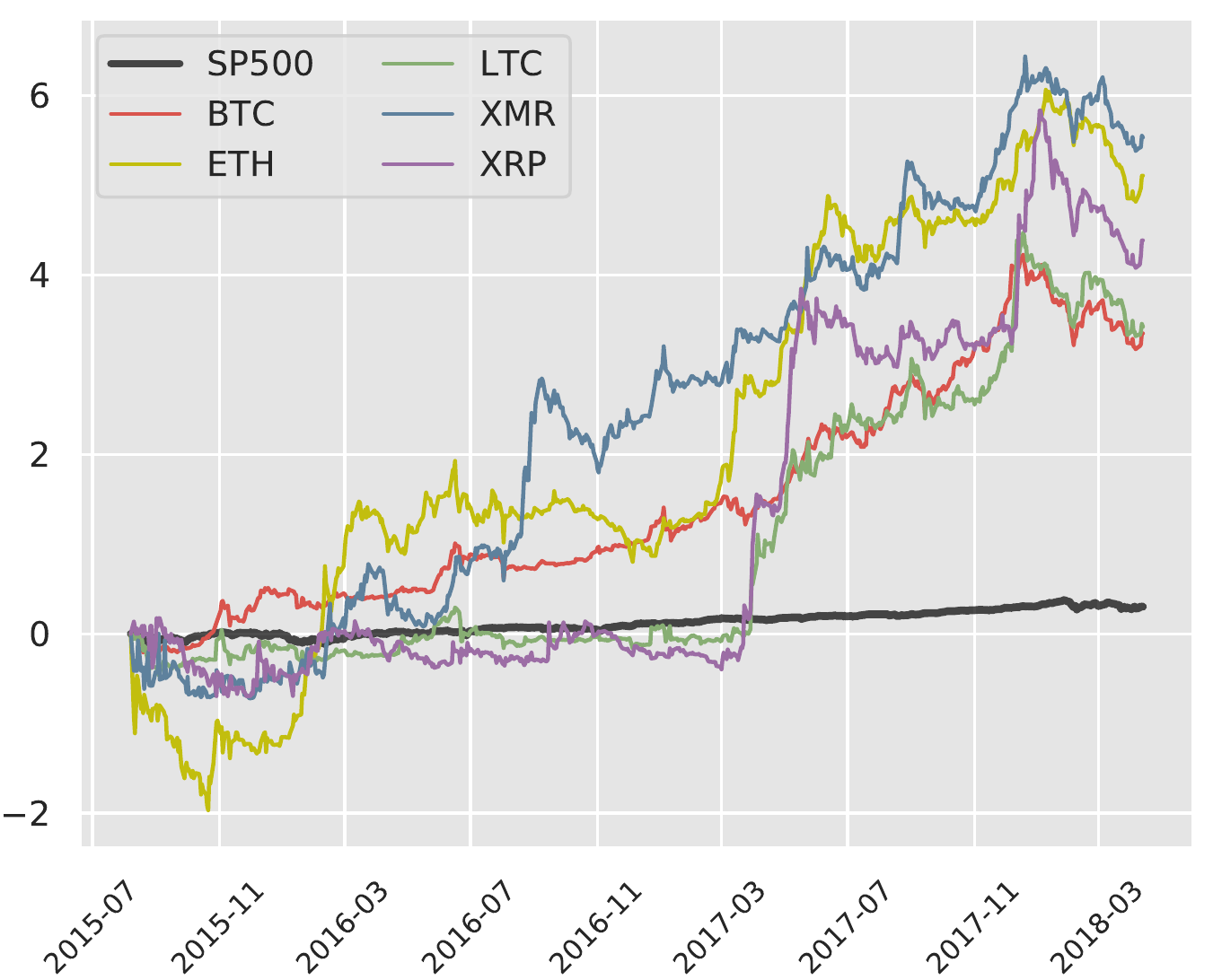}
\end{figure}

Figure \ref{fig:time_series} presents a time-series plot of growth rates computed as $\textrm{ln}(P_t/P_0)$ for the five cryptocurrencies, as well as the SP500 index for the purpose of comparison. Relative to the SP500 index, which has grown by over 30 percent during the August 2015 -- April 2018 period itself, price growth of digital assets seems incredible. The peak growth rate was reached by the most anonymous of the digital assets -- Monero (XMR) -- at the end of 2017. In fact, over the 28 month period to Dec 2017 all five cryptocurrencies exhibited returns in the range between 400 and 600 percent. Although the last four months of the sample period recorded significant drops in digital asset prices, the remaining growth amounts to more than a ten-fold return of the SP500 index.   

Table \ref{tbl:summary_stats} provides summary statistics for the daily log-return series of the thirteen variables considered here. As discussed Monero (XMR) exhibits the highest average daily return  of 0.82\%, followed by Etherium (ETH) at 0.76\%. Although these assets also exhibit high levels of risk, as measured by standard deviations in the second column, they still rank second on the risk adjusted (mean/standard deviation) basis. The best risk adjusted return is attained by Bitcoin (BTC), while the remaining cryptocurrencies come in the fourth and fifth place according to risk adjusted returns. Amongst the mainstream assets oil (WTI) has the highest mean return of 0.06\%, followed by SP500 and the broad commodities index SPGCSI with average daily returns of 0.04\%. All digital assets exhibit large extreme positive and negative returns as judged by the Min and Max columns. In fact, all thirteen variables fail the Jarque-Bera (JB) normality test, with JB statistics exceeding the 1\% critical value of 9.21.
Interestingly we observe that the geopolitical risk index (GPR) exhibits significant daily variability, with largest daily standard deviation, minimum and maximum daily changes. This reflects jumps present in the index series due to the sporadic nature of geopolitical incidents.      

\begin{table} [!h]
\begin{threeparttable}
\small
\captionsetup{justification=centering}
\caption{Daily Return Summary Statistics (7 August 2015 -- 13 April 2018)} \label{tbl:summary_stats} 
\begin{tabular}{lrrrrrrrr}
\toprule
{} &  Mean &   Std. &     Min &  Median &     Max &  JB Stat. &  Mean/Std. &  ADF Stat. \\
\midrule
BTC    &  0.50 &   4.76 &  -24.59 &    0.43 &   22.76 &    526.42 &       0.10 &     -26.15 \\
ETH    &  0.76 &   9.80 &  -91.63 &    0.13 &   49.76 &   5482.23 &       0.08 &     -26.57 \\
LTC    &  0.51 &   7.04 &  -31.25 &    0.00 &   55.16 &   6422.02 &       0.07 &     -23.27 \\
XMR    &  0.82 &  10.71 &  -51.08 &    0.00 &   75.05 &   2526.89 &       0.08 &     -22.09 \\
XRP    &  0.65 &  10.10 &  -56.33 &   -0.28 &   74.08 &   3250.14 &       0.06 &     -26.41 \\
SP500  &  0.04 &   0.85 &   -4.18 &    0.05 &    3.84 &    537.89 &       0.05 &     -26.25 \\
10YR   & -0.01 &   0.36 &   -1.89 &    0.01 &    1.46 &     59.28 &      -0.02 &     -27.60 \\
WTI    &  0.06 &   2.43 &   -8.08 &    0.10 &   11.29 &    149.85 &       0.03 &     -26.33 \\
GOLD   &  0.03 &   0.86 &   -3.38 &    0.02 &    4.59 &    259.10 &       0.04 &     -26.23 \\
USD    & -0.01 &   0.42 &   -2.40 &    0.00 &    2.49 &    355.51 &      -0.03 &     -25.81 \\
SPGSCI &  0.04 &   1.26 &   -4.49 &    0.06 &    5.26 &     46.22 &       0.03 &     -27.28 \\
TED    &  0.13 &   6.71 &  -34.17 &    0.00 &   27.87 &    206.14 &       0.02 &     -21.08 \\
GPR    &  0.27 &  70.17 & -268.40 &   -0.26 &  313.81 &     54.39 &       0.00 &     -19.79 \\
\bottomrule
\end{tabular}

\begin{tablenotes}
\setlength\labelsep{0pt}
\footnotesize
\item \textbf{Notes:} JB Stat. is the  \citeasnoun{jarque1987test} normality test statistic; JB critical values: 1\%: 9.21, 5\%: 5.99, 10\%: 0.21. ADF Stat. is the \citeasnoun{dickey1979distribution} unit-root test statistic; ADF critical values: 1\%: -3.44, 5\%: -2.87, 10\%: -2.60.
\end{tablenotes}
\end{threeparttable}
%\end{threeparttable}
\end{table}

Lastly, we consider the Augmented Dickey-Fuller (ADF) unit-root statistics provided in the last columns of the table. Since the ADF test is a left-sided test, all thirteen return series reject the null hypothesis of unit root at the 1\% significance level. This suggests that the daily returns series are stationary and suitable for further analysis using the methods outlined in the next section. 

\section{Methodology}
We base our analysis on a vector autoregressive model of order \textit{p}, denoted VAR(\textit{p}), which is specified for daily return series as follows:
\begin{eqnarray} \label{model_VAR}
y_t  & = & c+A_1y_{t-1}+A_2y_{t-2}+\dots+A_py_{t-p}+u_t.
\end{eqnarray}
In the above equation $y_t$ is a $(13\times1)$ vector containing the five crytocurrencies, six mainstream assets as well as the two risk factors discussed in Section \ref{section_data}. The current value of each of the thirteen variables can potentially depend on its own \textit{p} past values, as well as on \textit{p} lags of each of the other twelve variables. Parameter matrices $A_i$ are fixed and of dimension $13\times 13$ , while $c$ represents a $(13\times1)$ intercept vector. Finally, $u_t=(u_{1t},u_{2t},\dots,u_{13t})^{'}$ is a 13-dimensional innovation process characterized by the following properties: $\mathbb{E}(u_t)=0$, $\mathbb{E}(u_tu_t^{'})=\Sigma_u$ where $\Sigma_u$ is nonsingular, and $\mathbb{E}(u_tu_s^{'})=0$ for $s\neq t$. 

We employ two measures of connectedness derived from Granger-causality tests and forecast error variance decompositions as discussed in the following two subsections.

\subsection{Granger-Causality Networks} \label{section_Granger}

Empirical models testing Granger-causality relations specify equations where the dependent (LHS) variable $y$ is measured at time \textit{t}, while the explanatory (RHS) variables include lagged values of another variable $x$ recorded at $t-1, t-2,$ etc., as well as past values of $y$ itself. Testing whether $x$ Granger-causes $y$ then amounts to testing whether including the lagged values of $x$ can produce forecasts of $y$ which are superior to the predictions of $y$ constructed by relying only on the historical values of $y$ itself.
 
While the above discussion considers Granger causality within a bivariate system of $x$ and $y$, the idea may be extended to multivariate settings containing additional variables. In fact conducting the analysis while controlling for indirect sources of causality helps minimize the possibility of finding spurious relations. Spurious causality can occur in bivariate systems when a third variable, e.g. $z$, causes both $x$ and $y$ but is excluded from the analysis, see e.g. \citeasnoun{granger1969investigating} p. 429. Thus, in our investigation we employ the thirteen variable system (\ref{model_VAR}), rather than 78 corresponding bivariate models. 

On the other hand, a disadvantage of testing for Granger causality in a high-dimensional system is a complication regarding the inference about multi-step ($h$-step) causality discussed in \citeasnoun{luetkepohl2005new}, p. 49. To illustrate the issue consider a three-variable system where it is possible for $z$, for example, to be 1-step noncausal for $y$ while it is $h$--step causal when $h>1$. This eventuality arises due to the possibility that $z$ can cause $x$ at horizon 1 whereas $x$ is causal for $y$. Thus $x$ channels causality from $z$ to $y$ at intermediate forecast horizons. Testing for multi-step Granger causality is not straightforward, as discussed in \citeasnoun{luetkepohl1993testing}.

In order to eliminate potential complications discussed above we rely on the definition of causality originally proposed by \citeasnoun{granger1969investigating} that is based on 1-step ahead predictions. Within our $13$-dimensional VAR, we say that $y_j$ is Granger-causal for $y_i$ $(y_j{\rightarrow}y_i)$, for $i,j=1,\dots 13, i\neq j$, if the $1$-step ahead forecasts of $y_i$ made on the basis of the past values of the entire $(13\times1)$ vector $y$ are superior to the predictions of $y_i$ created when $y_j$ is excluded from the vector $y$. The above definition of causality is discussed within the VAR$(p)$ framework in \citeasnoun{dufour1998short}. In essence, testing for Granger noncausality involves testing relevant elements of the $A_1,\dots,A_p$ parameter matrices in (\ref{model_VAR}). Thus, we say that $y_j$ fails to Granger cause $y_i$ $(y_j\not\rightarrow y_i)$, in the sense of 1-step ahead predictions, if $A_{ij,\ell}=0$ for $\ell=1,\dots,p$, $i,j=1,\dots 13, i\neq j.$ These restrictions can be easily tested using standard Wald tests. 

After conducting a series of Granger-causality tests on the 13-dimensional system (\ref{model_VAR}) we follow \citeasnoun{billio2012econometric} and define network connections via indicator functions of the following form

$(y_j{\rightarrow}y_i)^q = \begin{cases} 1 \textrm{ if } y_j \textrm{ Granger causes } y_i \textrm{ at the } q \textrm{ percent level of significance}\\ 0 \textrm{ otherwise}
\end{cases}$

\subsection{Forecast Error Variance Decompositions Networks} \label{section_GVD}
Our second approach to measuring connectedness between digital assets relies on constructing a network using forecast error variance decompositions. The variance of the forecast error around the $h$-step forecast $\hat{y}_{t+h|t}$ constructed from (\ref{model_VAR}) is given by

\begin{equation} \label{eq_MSE}
\textrm{Var}(h)=\Sigma_u+\Theta_1\Sigma_u\Theta_1^{'}+\dots,+\Theta_{h-1}\Sigma_u\Theta_{h-1}^{'},
\end{equation}
where $\Sigma_u=\mathbb{E}(u_tu_t^{'})$, and $\Theta_i$'s represent the moving average representation coefficients for $i=1,\dots,h-1$. We also know that $\textrm{Var}(h)$ approaches $\textrm{Var}(y_t)$ for large forecast horizons $h$.

Forecast error variance decompositions $\textrm{VD}_{i,j}$ attempt to decompose the uncertainty $\textrm{Var}_i(h)$\footnote{These can be simply computed as $\textrm{Var}_i(h)=e_i^{'}\textrm{Var}(h)e_i$ where $e_i$ is the $i$th element selection vector.} about each asset $y_i$'s prediction $\hat{y}_{i, t+h|t}$ into fractions contributed by the shocks to every variable $y_j\in y$ for $i,j=1,\dots,13, i\ne j$. 

There are several ways to computing variance decompositions. We employ a technique known as the generalized forecast error variance decomposition proposed in \citeasnoun{pesaran1998generalized}. Unlike the orthogonalized variance decompositions, which typically depend on the ordering of the variables in the vector $y$ as discussed for e.g. in \citeasnoun{luetkepohl2005new} p. 64, generalized decompositions are order invariant. This is achieved by treating each of the variables as the first variable in the system. Generalized variance decompositions (GVD) computed from the $h$-step ahead forecasts are constructed as follows
\begin{equation}
\textrm{GVD}_{i,j}^h= \sigma_{jj}^{-1}\frac{[e_i^{'}(\Sigma_u+\Theta_1\Sigma_u+\dots,+\Theta_{h-1}\Sigma_u)e_j]^2}{\textrm{Var}_i(h)}. \label{eq:GDV}
\end{equation}
Since $\textrm{GVD}_{i,j}^h$ do not sum up to 1, they are standardized so that each row (summing across $j$) adds up to 100 as in \citeasnoun{diebold2014network}. Thus, we compute the fraction of variable $y_i$ forecast uncertainty accounted for by the shocks in variable $y_j\in y$ using standardized generalized forecast error variance decompositions (SGVD) as follows
\begin{equation}
\textrm{SGVD}_{i,j}^h= \frac{\textrm{GVD}_{i,j}^h}{\sum_{j=1}^K \textrm{GVD}_{i,j}^h}. \label{eq:SGDV}
\end{equation}

\subsection{Comparing the Two Types of Connectedness} \label{section_comparison}
The reason for choosing to employ two different modes of analysis is the difference in the information content provided by the two methods. 

Granger causality, as defined in Section \ref{section_Granger}, tells us if past values of variable $y_j$ help predict the current value of $y_i$, after controlling for the impact of other variables $y_k \in y$, $k\neq j$. Causal relationships uncovered using this method are thus: a) direct relationships -- they are computed while controlling for other variables included in the system, b) lagged relations which apply to 1-step ahead predictions, and c) they are based on tests of statistical significance and are not informative about the magnitude of the effect.

On the other hand, variance decompositions are functions of both the lagged linkages, through the $\Theta$ parameter matrices in (\ref{eq:GDV}), as well as contemporaneous correlations between the shocks contained in $\Sigma_u$. In bivariate models, \citeasnoun{luetkepohl2005new} p. 44 shows that if $y_j$ in Granger-noncausal for $y_i$ ($y_j\bm{\not\rightarrow}y_i)$ then the moving-average parameter matrices $\Theta$ in (\ref{eq_MSE}) contain zero elements such that $y_i$ does not depend on the shocks to $y_j$. If the shocks to the two variables are also contemporaneously uncorrelated then it is easy to see from (\ref{eq:GDV}) that $\textrm{GVD}_{i,j}$ will be zero. Thus, while the relationship between Granger Causality and GVD is not entirely straightforward, the impact of $y_j$ on $y_i$ in GVD is likely to be smaller when $y_j$ is Granger-noncausal for $y_i$. An advantage of GVD is that they can be used to judge the magnitude of a predictor's contribution to future uncertainty in the variable of interest. For example, if we find that shocks to $y_j$ account for 5 percent of future uncertainty in $y_i$, while shocks to $y_p$ contribute 20 percent of the uncertainty in $y_i$ we would then conclude that $y_p$ is more important than $y_j$ in transmitting uncertainty to $y_i$. A disadvantage of SGVD, as implemented here, relative to Granger-causality tests is that they are not based on tests of statistical significance.
\section{Empirical Results}

We start by providing the results of Granger causality tests mapped into a network in Figure \ref{fig:GC_entire_network}. The figure is created on the bases of significant causal relationships\footnote{A VAR(1) model is identified by information criteria and fitted to the data first.} at the 5 percent significance level, and augmented using edges which are significant at the 10 percent level.

\subsection{Granger Causality Links}
\begin{figure}[!h] 
\caption{Granger-Causal Network}\centering\label{fig:GC_entire_network}
%\begin{tabular}{cc}
\includegraphics[height=5in,width=6in]{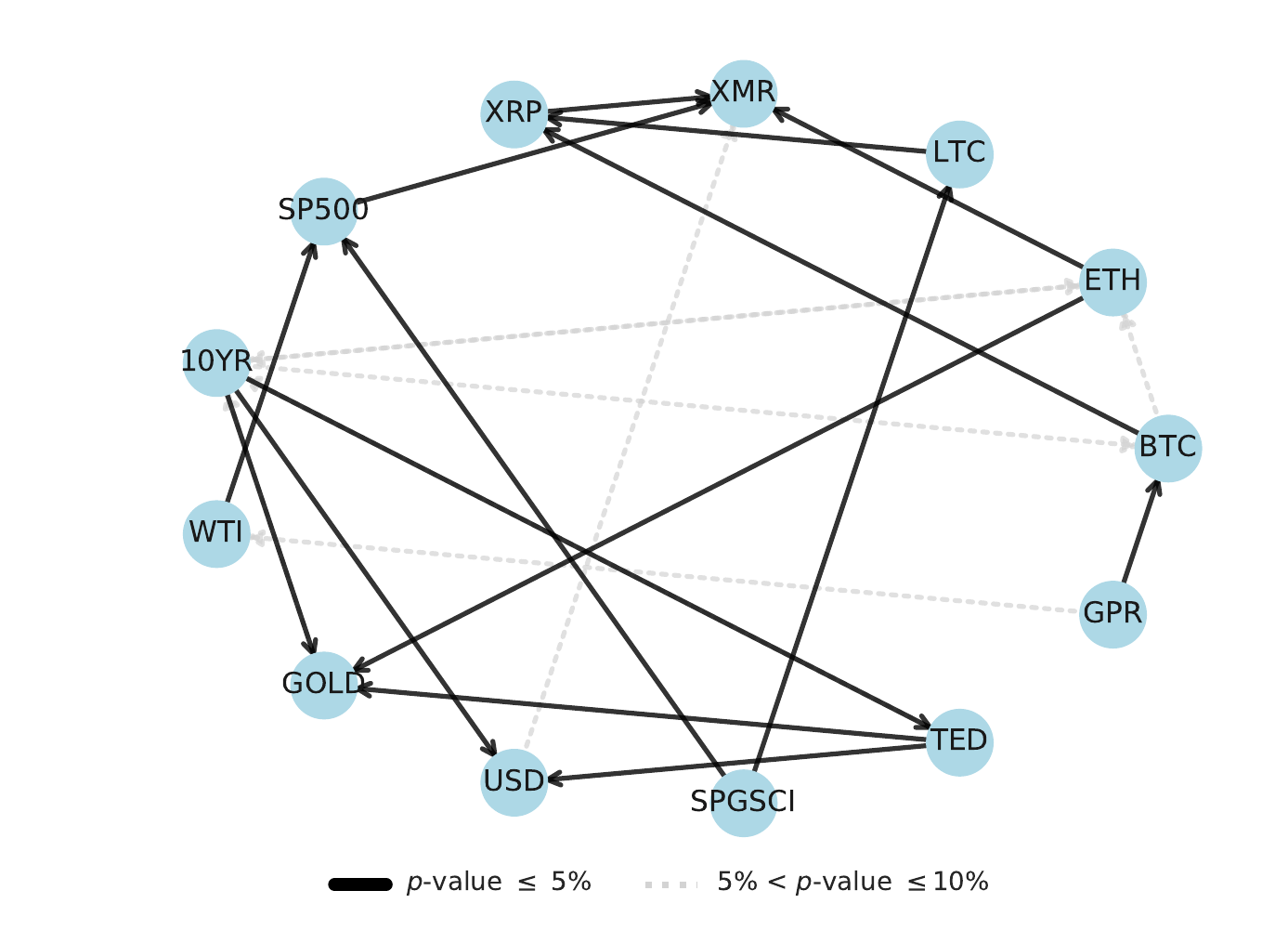}
%\end{tabular}
\end{figure}

Out of the total of 156 possible Granger-causal connections, we report 15 (23) statistically significant Granger-causal relationships at the 5 (10) percent significance. Of the 23 edges depicted in Figure \ref{fig:GC_entire_network}, five are estimated between digital assets and ten exist within the non-digital group. Considering cross-group linkages, we find two Granger-causal links from crypto assets to non-crypto group, and six relations in which non-crypto assets Granger cause digital assets. This suggests that mainstream assets in fact lead digital assets in more cases than the other way around. Table \ref{tbl:Granger_p_values} in the Appendix details $p$-values for all possible pairs.

According to Figure \ref{fig:GC_entire_network} the US 10 year bond index (10YR) is the most connected asset with a total of eight connections, followed by Etherium (ETH) which has five linkages and the TED spread with four direct Granger-causal relations. A number of these relations is bi-directional, such as for example the Granger-causal link between 10YR and ETH. These are signified by bi-directional arrows in the figure. 

Visually we observe that black lines, denoting Granger-causal links at the 5 percent significance, rarely cross the circle and mainly connect first or second neighbours. In contrast the grey lines are more represented in connecting assets across the digital and non-digital groups. 

Out of the total of 20 possible causal relations within the digital asset group, we find five statistically significant connections running in the following directions: Bitcoin (BTC) $\bm{\rightarrow}$ Ripple (XRP), BTC $\bm{\rightarrow}$ (ETH), ETH $\bm{\rightarrow}$ Monero (XMR), Lightcoin (LTC) $\bm{\rightarrow}$ XRP and XRP $\bm{\rightarrow}$ XMR. Interestingly, Ripple appears to be the most connected cryptocurrency within the crypto group with three connections. Within the non-digital group, the 10 year bond index (10YR) and the TED spread are each connected to three other variables. The following Granger-causal relationships are found: 10YR $\bm{\rightarrow}$ USD, USD $\bm{\rightarrow}$ 10YR, 10YR $\bm{\rightarrow}$ TED, Crude Oil (WTI)$\bm{\rightarrow}$Stocks (SP500), Commodities (SPGSCI)$\bm{\rightarrow}$SP500, TED$\bm{\rightarrow}$USD and Geopolitical Risk Index (GPR)$\bm{\rightarrow}$WTI.

Finally we consider Granger-causal relationships uncovered between the cryptocurrencies and the non-digital variables. In total there are eight statistically significant relationships, at the 10 percent level, out of the total of 80 potential links\footnote{Since there are 5 digital assets and 8 non-crypto variables, potentially there could $2\times5\times8=80$ Granger-causal relationships.}. The 10 year bond index appears to have three causal links with the cryptocurrencies. It causes price changes in Etherium and Bitcoin while it appears to be caused by Etherium (10YR$\bm{\leftrightarrow}$ETH,10YR$\bm{\rightarrow}$BTC). Bitcoin also seems to be caused by the Geopolitical Risk Index (GPR$\bm{\rightarrow}$BTC). Monero, which is the most anonymous of the digital assets considered here, is caused by SP500 (SP500$\bm{\rightarrow}$XMR).  We also observe commodities impacting Lightcoin (SPGSCI$\bm{\rightarrow}$LTC). Interestingly the USD only interacts with Monero (USD$\bm{\rightarrow}$XMR).

\subsection{Connectedness via Variance Decompositions}

\begin{figure}[!h] 
\caption{10-Day-Ahead Generalized Forecast Error Variance Decomposition Network}\centering\label{fig:GVD_entire_network}
%\begin{tabular}{cc}
\includegraphics[height=5in,width=6.5in]{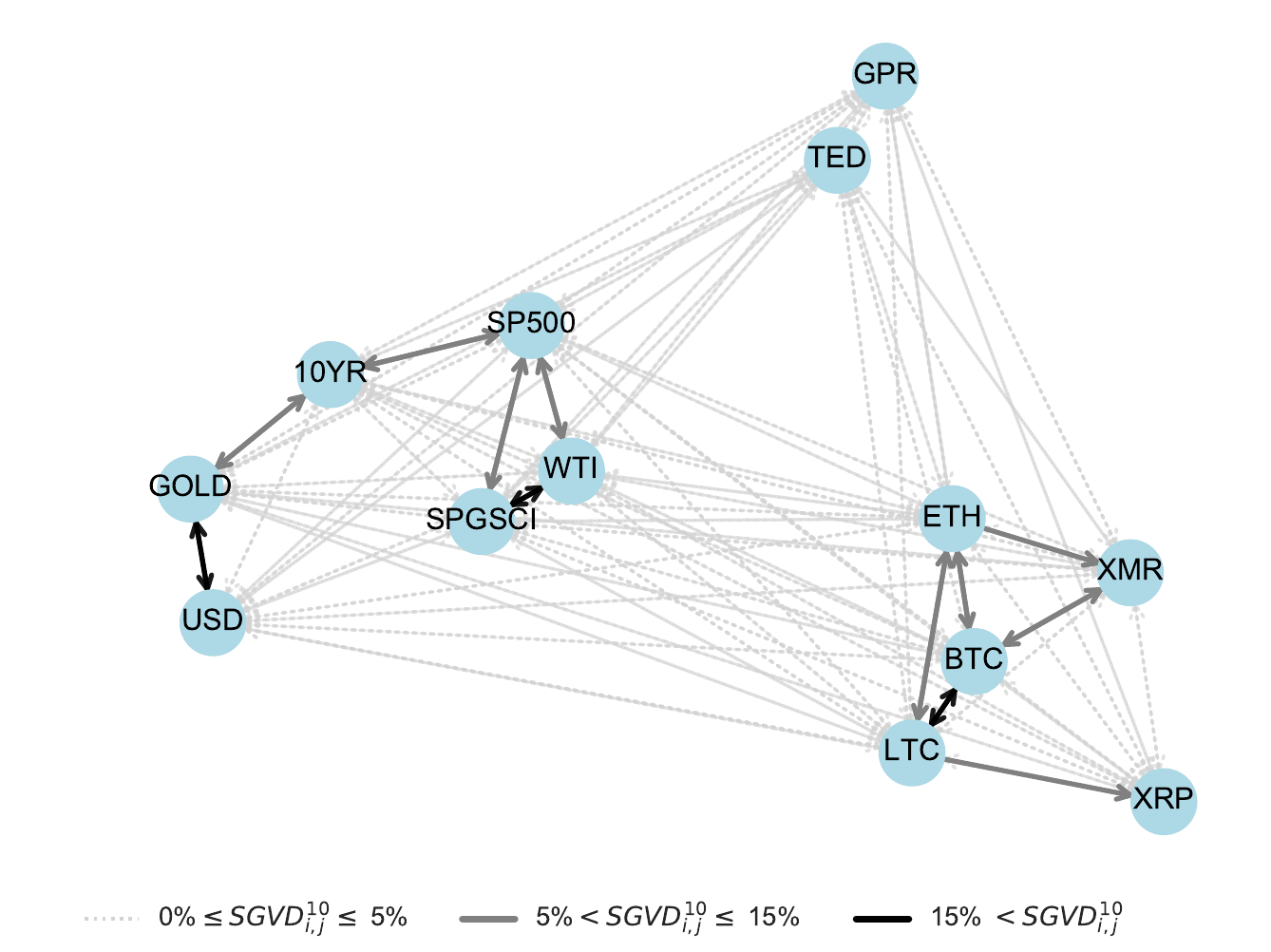}
%\end{tabular}
\end{figure}

Having explored and found eight statistically significant Granger-causal links across digital and non-digital asset groups, amongst other relations uncovered within the two groups, we now turn to map connectedness using (standardized) generalized forecast error variance decompositions (SGVD). These provide a measure of magnitude of each predictor's contribution to future uncertainty as discussed in Section \ref{section_comparison}. Table \ref{tbl:SGVD} provides the estimates of connectedness, which are also summarized visually in Figure \ref{fig:GVD_entire_network}.  The computed measures are based on 10-step ahead predictions, although 1-step and 20-step ahead SGVD yield similar results.

The first thing we notice from Table \ref{tbl:SGVD} is that the entries on the main diagonal are largest across every row. This implies that, as expected, own shocks contribute the most to each variable's future uncertainty. Second, we see that the table can be partitioned according to the magnitude of the elements into two large blocks -- one block containing the crypto currencies in the top-left corner, and the second block comprising non-crypto assets. 
In fact, we see that individual digital assets contribute less that 1 percent to any non-crypto variable (bottom-left block) and vice versa (top-right block). These groups can also be identified through the two columns and rows displayed in the margins of the table. For instance, Bitcoin receives 33.18 percent of its uncertainty from combined news shocks to the other four crypto assets while that figure is only 1.61 for the shocks sourced from non-crypto variables. Similarly, it contributes about 36.37 percent variability to other crypto currencies, and only 1.66 percent to non-crypto variables. Out of individual contributors to BTC's variability we see (across row one) that LTC accounts for about 18 percent while ETH and XMR rank second and third accounting for 6.7 and 6.07 percent of BTC's uncertainty.

Interestingly the USD does not contribute any significant variability to the cryptocurrencies, while it does to GOLD (about 17.45 percent) and the 10 year bond index (2.31 percent). Judging by this disconnectedness from the US dollar, it would appear that the cryptocurrencies have not yet developed as a competitor to the USD. 

In regards to the TED spread and the geopolitical risk index (GPR) we observe that they do not represent a significant source of risk to either crypto currencies or mainstream assets. The largest figure here is the 2.53 percent contribution of the 10 year bond return to the future uncertainty in the TED spread, which is a measure of credit risk.

\begin{landscape}
\begin{table} [!h]
\begin{threeparttable}
\small
\captionsetup{justification=centering}
\caption{Connectedness Table} \label{tbl:SGVD} 
\begin{tabular}{lrrrrrrrrrrrrrrrr}
\toprule
{} &    BTC &    ETH &    LTC &    XMR &    XRP &  SP500 &   10YR &    WTI &   GOLD &    USD &  SPGSCI &    TED & \multicolumn{2}{l}{GPR} &  $\genfrac{}{}{0pt}{}{From}{Cryptos}$ &  $\genfrac{}{}{0pt}{}{From}{Others}$ \\
\midrule
BTC                                &  65.21 &   6.70 &  17.67 &   6.07 &   2.74 &   0.42 &   0.15 &   0.12 &   0.04 &   0.23 &    0.11 &   0.06 &   0.47 &   &                                 33.18 &                                 1.61 \\
ETH                                &   8.00 &  77.07 &   6.83 &   4.98 &   1.59 &   0.04 &   0.43 &   0.26 &   0.14 &   0.01 &    0.13 &   0.01 &   0.51 &   &                                 21.40 &                                 1.52 \\
LTC                                &  17.78 &   5.83 &  66.44 &   3.94 &   4.57 &   0.34 &   0.17 &   0.05 &   0.09 &   0.15 &    0.23 &   0.38 &   0.03 &   &                                 32.11 &                                 1.45 \\
XMR                                &   7.03 &   5.20 &   4.59 &  77.42 &   3.61 &   0.71 &   0.08 &   0.44 &   0.06 &   0.17 &    0.25 &   0.22 &   0.22 &   &                                 20.44 &                                 2.14 \\
XRP                                &   3.56 &   1.64 &   6.57 &   3.42 &  83.84 &   0.20 &   0.15 &   0.08 &   0.22 &   0.08 &    0.06 &   0.01 &   0.16 &   &                                 15.19 &                                 0.97 \\
SP500                              &   0.51 &   0.21 &   0.38 &   0.18 &   0.19 &  69.29 &   8.95 &   8.07 &   2.31 &   0.37 &    9.22 &   0.20 &   0.11 &   &                                  1.48 &                                29.23 \\
10YR                               &   0.08 &   0.18 &   0.25 &   0.07 &   0.05 &   8.55 &  66.92 &   4.73 &  11.09 &   2.20 &    4.62 &   1.17 &   0.11 &   &                                  0.63 &                                32.45 \\
WTI                                &   0.05 &   0.02 &   0.08 &   0.12 &   0.11 &   5.76 &   3.41 &  49.30 &   0.01 &   0.92 &   39.99 &   0.02 &   0.21 &   &                                  0.38 &                                50.31 \\
GOLD                               &   0.03 &   0.48 &   0.08 &   0.03 &   0.08 &   2.13 &  11.44 &   0.10 &  66.41 &  17.45 &    0.44 &   1.27 &   0.06 &   &                                  0.70 &                                32.89 \\
USD                                &   0.18 &   0.04 &   0.06 &   0.07 &   0.00 &   0.23 &   3.18 &   1.60 &  18.98 &  71.64 &    2.58 &   1.17 &   0.27 &   &                                  0.36 &                                28.00 \\
SPGSCI                             &   0.15 &   0.00 &   0.18 &   0.03 &   0.10 &   6.44 &   3.21 &  39.29 &   0.19 &   1.59 &   48.52 &   0.16 &   0.14 &   &                                  0.46 &                                51.03 \\
TED                                &   0.36 &   0.23 &   0.53 &   0.41 &   0.08 &   0.52 &   2.53 &   0.04 &   0.65 &   1.12 &    0.01 &  93.19 &   0.33 &   &                                  1.60 &                                 5.21 \\
GPR                                &   0.29 &   0.68 &   0.05 &   0.41 &   0.23 &   0.31 &   0.15 &   0.22 &   0.03 &   0.26 &    0.08 &   0.09 &  97.21 &   &                                  1.65 &                                 1.13 \\
                                   &        &        &        &        &        &        &        &        &        &        &         &        &        &   &                                       &                                      \\
$\genfrac{}{}{0pt}{}{To}{Cryptos}$ &  36.37 &  19.37 &  35.67 &  18.41 &  12.51 &   1.71 &   0.98 &   0.96 &   0.54 &   0.65 &    0.79 &   0.67 &   1.39 &   &                                       &                                      \\
$\genfrac{}{}{0pt}{}{To}{Others}$  &   1.66 &   1.84 &   1.61 &   1.31 &   0.84 &  23.93 &  32.87 &  54.05 &  33.25 &  23.92 &   56.92 &   4.08 &   1.24 &   &                                       &                                      \\
\bottomrule
\end{tabular}

\begin{tablenotes}
\setlength\labelsep{0pt}
\footnotesize
\item \textbf{Notes:} The \textit{ij}th entry gives the percent of 10-day ahead forecast uncertainty of $y_i$ due to shocks to $y_j$. The measures are computed using eq. (\ref{eq:SGDV}).
\end{tablenotes}
\end{threeparttable}
%\end{threeparttable}
\end{table}
\end{landscape}

Figure \ref{fig:GVD_entire_network} summarizes the connectedness relations discussed above. Here the nodes are connected with the edges corresponding to the estimates of $\textrm{SGVD}_{i,j}^{(10)}$, that is fractions of each return $y_i$'s 10-step ahead forecast error variance accounted for by the shocks to the variable $y_j$. Linestyles and color schemes are used to denote different threshold levels which are set to 5 and 15 percent. The reader can refer to Table \ref{tbl:SGVD} for the exact magnitudes of the depicted connections.

As explained above we observe three main groupings of variables. Digital currencies are linked together in the bottom-right portion of the network graph with relatively strong linkages, the set of mainstream assets is in the middle while the two risk factors are at the top of the network graph.

Amongst the cryptocurrencies, we see that most of the solid grey lines are bi-directional with arrows on each side. It appears that news shocks to Etherium (ETH) moderately impact the uncertainty in Monero (XMR), Bitcoin (BTC) and Lightcoin (LTC). On the other hand, the impacts of BTC and LTC on ETH predictions are in a similar range, while XMR effect is weaker. The strongest relationship appears to be between BTC and LTC exceeding the 15 percent threshold. This link also appears to be bi-directional. Lastly, LTC affects Ripple (XRP) with a moderate magnitude.

Considering the connectedness between the set of mainstream assets we see that the strongest links are between the USD and GOLD, as well as between commodities index SPGSCI and oil (WTI). All of these connections exceed 15 percent of the generalized variance decomposition and are bi-directions. SP500 stock index is moderately connected with the 10 year bond index, while WTI and SPGSCI are linked via bi-directional connections. Lastly, GOLD and 10YR mutually contribute between 5 and 15 percent of forecast error variances.    

Lastly, the network graph clearly illustrates the lack of substantial links across the digital and non-digital asset groups. These results are largely in line with the findings of \citeasnoun{corbet2018exploring}.
\section{Conclusions}
We explore connectedness within and across digital and non-digital asset groups, as well as two risk factors comprising the TED spread and a geopolitical risk index. Our connectedness measure drown from the concept of Granger causality is based on tests of statistical significance, and as such it identifies linkages which are unlikely to be due to random chance. However, it are largely uninformative about the magnitude of the identified links. Therefore we supplement the analysis with forecast error variance decompositions to provide a measure of relative magnitudes of connectedness. 

Granger-causal network suggests that most of the statistically significant linkages are within the groups of digital currencies and mainstream assets. However, a number of digital/non-digital asset linkages are found too. Out of the total of 80 cross-pairs we find 6 statistically significant Granger-causal relations from non-digital to digital assets, and 2 causations from digital to mainstream assets. For instance, the USD is found to impact Monero, while SPGSCI commodity index Granger causes price changes in Lightcoin. Bitcoin and Etherium appear to be caused by changes in the 10 year government bond index. Of the two risk factors, the geopolitical risk index Granger causes Bitcoin at the 5 percent level of significance. Standardized generalized error variance decompositions (SGVD) are depicted in Figure \ref{fig:GVD_entire_network}, which illustrates relative weakness of the uncovered connections between digital and non-digital assets. The exact magnitudes are reported in Table \ref{tbl:SGVD}, and show that none of the individual across-group asset links is greater than 1 percent. 

In summary, while there are some time-series links between digital and mainstream asset classes these connections are of relatively small magnitude. From a systemic risk point of view this implies that the risk of price distress being transmitted from crypto currencies to non-digital asset is small. From a diversification point of view, one may argue in favour of investing in digital currencies as they appear to be relatively insulated from market trends found in major asset classes.

\newpage
\bibliography{CryptoSVAR}

\begin{landscape}
\section{Appendix}
\setcounter{table}{0}
\renewcommand{\thetable}{A\arabic{table}}

\begin{table} [!h]
%\centering
\begin{threeparttable}
\caption{Granger Causality p-values} \label{tbl:Granger_p_values} \bigskip
\begin{tabular}{lrrrrrrrrrrrrr}
\toprule
Causality From $\rightarrow$  &    BTC &    ETH &    LTC &    XMR &    XRP &  SP500 &   10YR &    WTI &   GOLD &    USD &  SPGSCI &    TED &   GPRD \\
\midrule
BTC    &        &  0.308 &  0.344 &  0.697 &  0.644 &  0.438 &  0.080 &  0.164 &  0.588 &  0.241 &   0.183 &  0.954 &  0.049 \\
ETH    &  0.081 &        &  0.381 &  0.648 &  0.148 &  0.633 &  0.089 &  0.377 &  0.986 &  0.662 &   0.585 &  0.699 &  0.219 \\
LTC    &  0.231 &  0.418 &        &  0.973 &  0.220 &  0.926 &  0.200 &  0.102 &  0.750 &  0.187 &   0.024 &  0.654 &  0.554 \\
XMR    &  0.758 &  0.013 &  0.239 &        &  0.037 &  0.027 &  0.929 &  0.610 &  0.107 &  0.089 &   0.819 &  0.918 &  0.894 \\
XRP    &  0.013 &  0.576 &  0.000 &  0.711 &        &  0.538 &  0.515 &  0.795 &  0.585 &  0.655 &   0.916 &  0.883 &  0.640 \\
SP500  &  0.454 &  0.112 &  0.751 &  0.111 &  0.618 &        &  0.556 &  0.030 &  0.891 &  0.161 &   0.021 &  0.761 &  0.634 \\
10YR   &  0.532 &  0.077 &  0.220 &  0.366 &  0.780 &  0.838 &        &  0.680 &  0.056 &  0.197 &   0.429 &  0.066 &  0.406 \\
WTI    &  0.754 &  0.733 &  0.407 &  0.590 &  0.241 &  0.496 &  0.725 &        &  0.554 &  0.407 &   0.648 &  0.538 &  0.095 \\
GOLD   &  0.929 &  0.048 &  0.902 &  0.865 &  0.853 &  0.728 &  0.007 &  0.633 &        &  0.654 &   0.197 &  0.003 &  0.408 \\
USD    &  0.185 &  0.425 &  0.689 &  0.973 &  0.894 &  0.547 &  0.011 &  0.267 &  0.676 &        &   0.854 &  0.016 &  0.347 \\
SPGSCI &  0.661 &  0.482 &  0.388 &  0.816 &  0.410 &  0.251 &  0.763 &  0.198 &  0.696 &  0.761 &         &  0.135 &  0.171 \\
TED    &  0.254 &  0.414 &  0.621 &  0.576 &  0.714 &  0.571 &  0.030 &  0.452 &  0.557 &  0.132 &   0.388 &        &  0.198 \\
GPRD   &  0.354 &  0.381 &  0.150 &  0.938 &  0.738 &  0.328 &  0.791 &  0.314 &  0.780 &  0.666 &   0.703 &  0.487 &        \\
\bottomrule
\end{tabular}
\end{threeparttable}
\end{table}

\end{landscape}

\end{document}